\documentclass[conference]{IEEEtran}
\IEEEoverridecommandlockouts
\usepackage{cite}
\usepackage{amsmath,amssymb,amsfonts}
\usepackage{algorithmic}
\usepackage{graphicx}
\usepackage{textcomp}
\usepackage{xcolor}
\usepackage{float}

\def\BibTeX{{\rm B\kern-.05em{\sc i\kern-.025em b}\kern-.08em
    T\kern-.1667em\lower.7ex\hbox{E}\kern-.125emX}}
\begin{document}

\title{AIOptimizer - Software performance optimisation prototype
for cost minimisation\\

}

\author{\IEEEauthorblockN{Noopur Zambare}
\IEEEauthorblockA{\textit{Department of Mechanical Engineering} \\
\textit{Indian Institute of Technology}\\
Jodhpur, India \\
zambare.1@iitj.ac.in}

}

\maketitle

\begin{abstract}
This study presents AIOptimizer, a prototype for a cost-reduction-based software performance optimisation tool. The study focuses on the design elements of AIOptimizer, including user-friendliness, scalability, accuracy, and adaptability. To deliver efficient and user-focused performance optimisation solutions, it promotes the use of robust integration, continuous learning, modular design, and data collection methods. The paper also looks into AIOptimizer features including collaboration, efficiency prediction, cost optimisation suggestions, and fault diagnosis. Additionally, it introduces AIOptimizer, a recommendation engine for cost optimisation based on reinforcement learning, and examines several software development life cycle models. The goal of this research study is to showcase AIOptimizer as a prototype that continuously improves software performance and reduces costs by utilising sophisticated optimisation techniques and intelligent recommendation systems. Numerous software development life cycle models, including the Big Bang, V-, Waterfall, Iterative, and Agile models are the subject of the study. Every model has benefits and drawbacks, and the features and requirements of the project will decide how useful each is.

\end{abstract}


\section{Introduction}
AIOptimizer is a cost-optimized software performance optimisation tool. It has a recommendation system based on reinforcement learning to increase the affordability and effectiveness of software systems. 
It provides effective and user-centric performance optimisation solutions by utilising a modular design, data-gathering techniques and reliable integration. It prioritises security and privacy while also providing defect discovery, cost optimisation advice, performance prediction, and collaborative features.\\

In addition, the article explores alternative SDLC models and introduces AIOptimizer's reinforcement learning-based system of recommendations for cost optimisation. The software development life cycle (SDLC) is a methodical approach to software development that specifies the phases involved in the creation and maintenance of software systems. It includes everything from strategy and gathering requirements to programming, testing, deployment, and maintenance. The SDLC is a systematic framework that ensures software development projects are organised, efficient, and consistent with client expectations. Software development teams can successfully manage resources, eliminate risks, and produce quality software products on schedule and within budget by adhering to the SDLC.
The model used here is the Spiral Model for the Software Development Life Cycle. Its overall goal is to continuously enhance software performance and save costs by employing advanced optimisation techniques and smart recommendation systems.\\

Furthermore, the paper describes the AIOptimizer recommendation system for cost optimisation, which employs reinforcement learning. It is able to offer smart recommendations on the basis of observable system statuses and user feedback thanks to this approach. AIOptimizer continually enhances its cost optimisation algorithms by applying reinforcement learning techniques, contributing to improved software performance and cost savings.

\section{Methodology}
\subsection{Crucial factors of designing}
Important design considerations for AIOptimizer include ensuring the accuracy and dependability of optimisation recommendations, giving adaptability and versatility to diverse software systems, maintaining a user-friendly interface, designing for scalability and efficiency, placing a premium on privacy and security of data, and enabling continuous learning and improvement. AIOptimizer should be able to provide efficient and user-centric software performance and cost optimisation solutions by considering these factors.  \newline
        
                \subsubsection{Data Collection and Monitoring \vspace{5pt}\newline 
                    Determine the sources of needed information for AIOptimizer. This could involve log files, monitoring of performance tools, programme metrics, architecture data, configuration settings, and any other data that can shed light on software efficiency and cost-related factors. \vspace{5pt}\newline
                    Implement data collection mechanisms from the specified sources. This may involve merging current monitoring systems, utilising APIs, collecting logs, or developing custom data extraction methods. Confirm that the data acquisition mechanisms are effective, trustworthy, and capable of gathering the required data in a timely manner. Determine the appropriate granularity level for the gathered data. Data can be collected at various levels, including per business, per user, per component, or even at an aggregated system level, depending on the optimisation objectives and system requirements.}

                \subsubsection{Modularity and Componentization \vspace{5pt}\newline
                Modularity and componentization are crucial design concepts for AIOptimizer because they promote adaptability, maintainability, and scalability. Using a modular architecture that separates the system into discrete, independent modules or components. Each module concentrates on particular functionalities which make the system simpler to comprehend, develop, test, and manage.
                   \vspace{5pt}\newline
                  Components should be designed to be reusable, meaning that they can be readily utilised in various components of a system or in different projects. By reusing existing components, reusable lowers redundant work, encourages code collaboration, and accelerates development.
}

                \subsubsection{Feedback Loop and Continuous Learning \vspace{5pt}\newline
                Its essential components are the feedback loop and perpetual learning, which allow the system to develop and adapt over time. Using a reward function, reinforcement learning achieves its feedback cycle. AIOptimizer specifies a reward function that measures the performance enhancement and cost savings resulting from specified actions. The reward function gives a feedback signal that directs the reinforcement model for learning towards more beneficial actions. \vspace{5pt}\newline
                It continuously trains and modifies its reinforcement learning algorithm based on the monitored states, actions taken, and rewards received. The model improves its decision-making abilities over time by learning from past data, including feedback and efficiency metrics.}

                \subsubsection{Integration and Scalability \vspace{5pt}\newline
                This integration may include APIs, database connectors, or interfaces to facilitate the exchange of information and compatibility with tracking tools, management of configuration systems, and other applicable software components. It should utilise distributed computing, parallel computation, or cloud-based resources to effectively manage large volumes of data and execute complex analytic and optimisation tasks.
                \vspace{5pt}\newline
                It must support both vertical and horizontal scalability to accommodate growing data volumes and workloads. Horizontal scalability enables the workload to be distributed across multiple machines whereas vertical scalability increases the CPU and memory on a single machine.}

\subsection{Working of AIOptimizer}
AIOptimizer optimises software performance and reduces costs by utilising novel methods such as reinforcement learning. It gathers performance metrics, resource utilisation, and cost factor data, that is utilised to train models and create optimisation recommendations. The system analyses the data, recognises patterns, and discovers optimal cost-reduction strategies. Through a user-friendly interface, users can interact with AIOptimizer to receive recommendations, customise settings, and collaborate with others. AIOptimizer continually learns from user feedback and adapts its models to evolving software environments, thereby assuring continuous performance optimisation and cost reduction enhancements.

        \subsubsection{Flaw Detection and Diagnosis}
        \begin{itemize}
        \setlength{\itemsep}{0pt}
            \item In software, flaw recognition and evaluation include recognising and diagnosing issues or faults inside the software system that may affect its performance or lead to increased costs. On the specified features, the software employs statistical evaluation and methods for detecting anomalies to identify irregular or unexpected patterns. Statistical methods, such as the median, standard deviation, mean, and percentile analysis, can assist in identifying deviations from expected behaviour. 
            \item It establishes performance metric thresholds or baselines according to historical data or preset expectations. These thresholds define permissible levels or limits for indicators of performance, and deviations outside of these thresholds may indicate potential defects. It employs rule-based analysis to identify known defects or patterns recognised through expertise in the field or prior experiences. These norms can be predefined or discovered through past data and expertise.
            \item It creates diagnostic insights and summaries that summarise the detected defects, their possible impact on efficiency and costs, and suggested corrective measures. The insights provide users with a comprehensive comprehension of the identified defects and their repercussions.
        \end{itemize}
        
        \subsubsection{Cost Optimization Recommendations}
        \begin{itemize}
        \setlength{\itemsep}{0pt}
            \item The software should collect data on cost metrics, resource utilisation, and other significant variables influencing software system expenses. The collated data is analysed to identify system-wide patterns, trends, and cost drivers. Software should identify and classify various cost components, including infrastructure costs, licencing fees, maintenance expenses, and operational costs.
            This classification facilitates comprehension of the various areas where cost optimisation can be targeted.
            \item On the basis of gathered data and user-defined objectives, software establishes cost optimisation objectives. These goals may include lowering infrastructure costs, optimising resource utilisation, minimising licencing fees, or enhancing operational efficacy. Based on the observed cost drivers and optimisation objectives, the software produces recommendations for cost optimisation. These recommendations may include actions such as optimising database queries, instituting caching mechanisms, and automating manual processes.
        \end{itemize}
        
        \subsubsection{Performance Prediction and Simulation}
        \begin{itemize}
        \setlength{\itemsep}{0pt}
            \item Cost-benefit analyses are performed for each recommended optimisation strategy by software. It takes into account possible cost savings, approximated implementation effort, and anticipated effect on system performance or other pertinent factors. This analysis assists in ranking the recommendations according to their prospective return on investment. This simulation aids in evaluating anticipated results and allows users to make sound choices regarding the prioritisation and practicability of the recommendations.
        \end{itemize}
        
        \subsubsection{Reporting and Visualization}
        \begin{itemize}
        \setlength{\itemsep}{0pt}
            \item In software, reporting and visualisation entail presenting pertinent insights, analytical results, and performance metrics in an understandable and graphical format. This makes it easier for users to comprehend and analyse the findings. 
            \item It establishes the key performance indicators (KPIs) pertinent to the performance and cost optimisation of the software system. KPIs can include response time, throughput, resource utilisation, cost per transaction, or any other particular metric aligned with optimisation objectives. It generates a dashboard that functions as the central interface for displaying generated reports and visualisations. The dashboard offers a summary of the application system's efficiency, cost status, optimisation progress, and important findings.
            
        \end{itemize}
        
        \subsubsection{Integration and Collaboration}
        \begin{itemize}
        \setlength{\itemsep}{0pt}
            \item It permits customers to export results and visualisations in a variety of formats, including PDF, Excel, and CSV. This makes it easier to share insights, findings, and suggestions with stakeholders, teammates, and management. It includes collaboration capabilities that promote communication, sharing of information, and teamwork. These features may include forums for discussion, remark sections, document sharing, and capabilities for real-time collaboration.
            \item APIs facilitate data exchange, system coordination, and process automation between software along with other suitable systems or applications. It offers team workspaces or project administration functionality in which stakeholders and team members can collaborate, designate tasks, and monitor progress. Workspaces serve as an administrative centre for exchanging data, discussing optimisation methods, and coordinating efforts.
            \item It integrates with version control systems such as Git to monitor changes, manage code repositories, and facilitate development team collaboration. This integration assures the correct versioning and synchronisation of code and optimization-related artefacts.
        \end{itemize}
    
\subsection{Software Development Life Cycle}
\begin{figure}[H]
      \centering
      \includegraphics[width=3.5in]{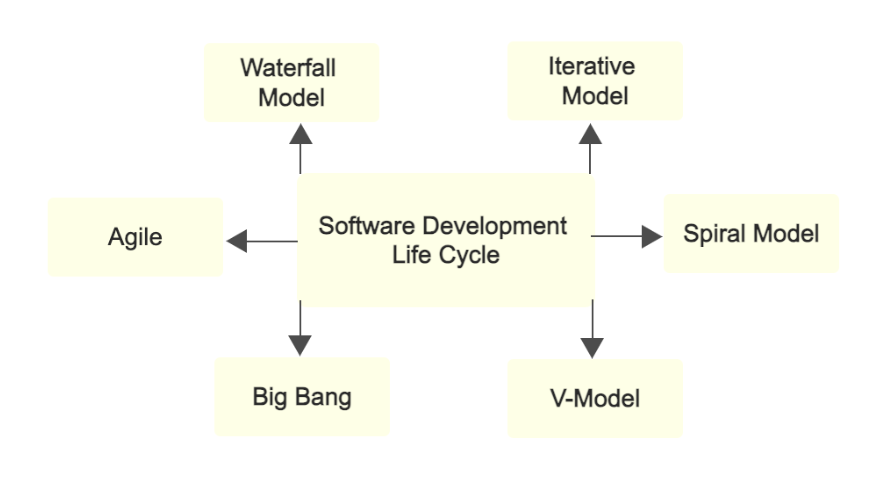}
      \caption{Software Development Life Cycle}
    \end{figure}
        \subsubsection{Waterfall Model}
        The Waterfall model is a linear, sequential and structured approach to software development. It consists of different stages that are implemented sequentially, with each phase's outputs building upon the previous phase's. It is appropriate for tasks with well-defined and stable requirements and where minimal changes are anticipated. However, it can cause difficulties when requirements are insufficiently understood or subject to frequent modification. 

        \subsubsection{Iterative Model}
        It incorporates aspects derived from the Waterfall model with repeated practises. It entails iterative cycles of gathering requirements, development, and testing in fewer stages. The advantages of the iterative model include adaptability, prompt feedback, and the capacity to accommodate altering requirements. It improves stakeholder engagement and increases the likelihood of delivering a refined and unified product. However, efficient task management and coordination are required to manage iterative cycles and guarantee continuous progress.

        \subsubsection{Spiral Model}
        Combining elements of both the Waterfall model and iterative development, the Spiral framework is a risk-driven software development process model. During the software development life cycle, it emphasises risk analysis and mitigation. The Spiral model is adaptable and iterative, enabling uninterrupted risk evaluation and reduction all over the software development process. It encourages active stakeholder participation, early risk identification and resolution, and adaptability to shifting requirements and priorities. To ensure successful implementation, constant tracking of risks and effective collaboration among project stakeholders are necessary. This model is best suited for AIOptimizer.

        \subsubsection{V-Model}
        The V model is a life cycle framework for software development that emphasises the connection between testing and development operations. It is frequently viewed as a modification of the Waterfall paradigm, in which testing activities are incorporated at every step of the development process. The V-Model employs a sequential methodology. However, its adaptability to changing requirements may be limited, and alterations in design or performance may necessitate adaptations across multiple stages.

        \subsubsection{Big Bang Model}
        The Big Bang framework is an unstructured approach to software development that lacks a defined life cycle. It is typically employed for small initiatives or software development experiments. Within the Big Bang model, there is no established phase or process sequence. Instead, development activities occur concurrently or in a manner that is not linear. Due to the lack of structure, scheduling, and formal procedures, the Big Bang model shouldn't be used for complex or substantial software projects. However, it may be suitable for certain scenarios.

        \subsubsection{Agile Model}
        Agile is a sequential and progressive software development methodology that emphasises adaptability, collaboration, and rapid software delivery. It is founded on the Agile Manifesto's guiding principles. Agile encourages flexible planning, growth through evolution, prompt and continuous delivery, and close communication among multidisciplinary groups and their stakeholders.
\subsection{Phases of Spiral Model}
\subsubsection{Planning}
        During this phase, optimisation objectives, measurements of performance, cost factors, and the software system to be optimised are specified.
        In addition to establishing the project schedule and allocating resources, the planning phase involves determining key users and their respective responsibilities.

        \subsubsection{Risk Analysis}
        This involves analysing technical hazards, limitations in data availability, security flaws, and any other possible challenges. Strategies for risk mitigation and backup plans are developed in order to deal with identified risks and lessen their impact on the undertaking.
        \begin{figure}[H]
          \centering
          \includegraphics[width=3.5in ]{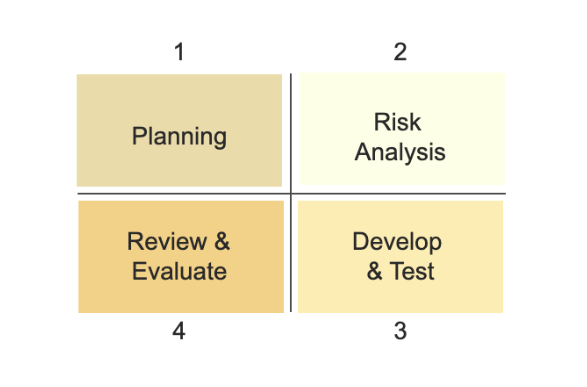}
          \caption{Phases of Spiral Model}
        \end{figure}
        \subsubsection{Develop and test}
        The developed components undergo extensive testing to ensure their precision, dependability, and compatibility with a variety of software systems and databases.

        \subsubsection{Review and evaluate}
        This phase includes a review of the implemented functionality, an evaluation of AIOptimizer's performance, and the collection of user feedback. The achievement of performance metrics, optimisation results, and cost savings is compared to the predetermined objectives and requirements.
        User acceptability testing and feedback from users play an essential part in evaluating AIOptimizer's usability, efficiency, and overall customer satisfaction. On the basis of the feedback and review, the software's capabilities are adjusted, refined, or enhanced as required.

\section{Recommendation System Using Reinforcement Learning for Cost Optimization}
Due to its simplicity of execution, distinct action space, flexibility, generalisation, interpretability and explain--
ability, progressive learning and continuous improvement, Q-learning is a suitable technique for the AIOptimizer
recommendation system. By integrating Q-learning into the recommendation system of AIOptimizer, you can
provide users with helpful information regarding the selection of optimisation techniques based on the observed
system state. The Q-learning algorithm utilises past information as well as user feedback to make intelligent rec-
ommendations, thereby contributing to the continual enhancement and optimisation of application performance
and cost reduction.
\begin{figure}[H]
  \centering
  \includegraphics[width=3.5in ]{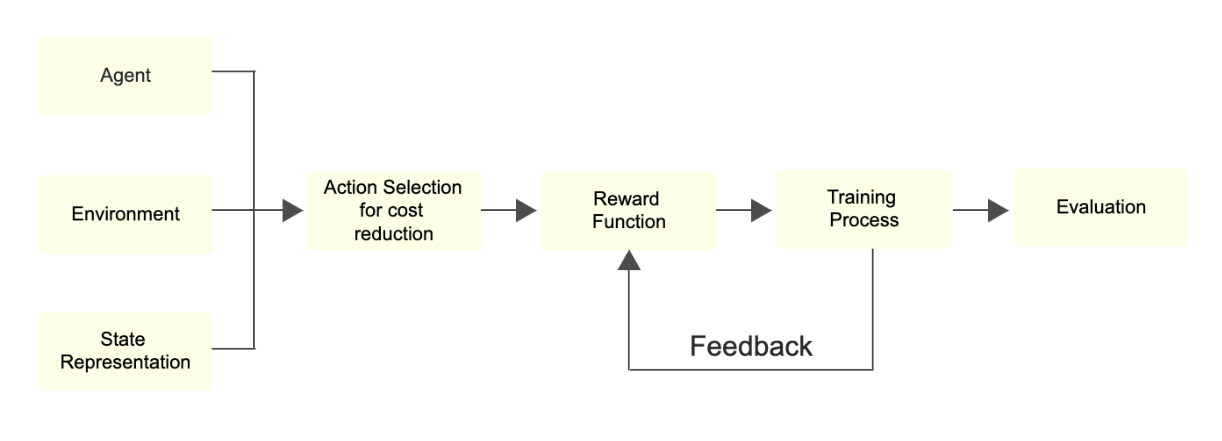}
  \caption{Design for Recommendation System}
\end{figure}
\subsection{Agent}
The agent in AIOptimizer is the recommendation algorithm responsible for optimising cost in the software system by choosing actions and taking action. It employs Q-learning to discover ideal cost-reduction methods based on recorded states and rewards. The agent engages with the environment and gets rewards as feedback, which guides its decision-making.

\subsection{Environment}
    The environment is the software system for which cost reductions are sought.
    It consists of all the system elements, configurations, and resources that influence cost. By choosing actions and observing the resultant states and rewards, the agent engages with the environment.

    \subsection{State Representation}
    The state reflects the current configuration or state of the software system, such as its resource allocation, performance metrics, and any other pertinent cost-influencing factors. The state representation should collect the information that allows the agent to make sound choices regarding cost reduction.
It could include metrics such as resource utilisation, speed of response, task distribution, or any additional cost-optimization-related parameters.

    \subsection{Action Selection for cost reduction}
    Depending on the observed state, the agent selects the most suitable steps to optimise cost as part of the action selection procedure. In regard to cost reduction, actions are different techniques that can result in cost reductions, such as resource scaling, workload balancing, and code optimisations. The agent makes informed decisions based on the acquired Q-values associated with every action-state couple, selecting actions that are anticipated to result in the greatest cost reduction.

    \subsection{Reward Function}
    The reward function incentivizes the agent to choose actions that result in significant cost savings while weighing the costs and benefits of other performance metrics. It can be intended to provide greater rewards for actions that result in significant cost reductions or to impose penalties for actions that result in cost increases.

    \subsection{Training Process}
    The agent learns optimal cost-reduction strategies by repeatedly engaging with the environment during the training process. During training, the agent observes the present state, chooses an action according to the exploration-exploitation trade-off, and then executes the action in the environment. After receiving feedback in the form of rewards, the Q-values in the Q-table are updated. The agent's policy progressively improves as it gains knowledge from the observed rewards, and the Q-values are updated accordingly.

    \subsection{Evaluation}
    Comparing the cost reduction attained through the agent's recommendations to baseline strategies serves as a metric for measuring the agent's effectiveness.
    To validate the cost-reduction capabilities of AIOptimizer, evaluation may entail modelling different situations or conducting experiments.

\section{Conclusion}
The software performance optimisation approach proposed in this research article, AIOptimizer, is intended to be useful in a wide range of target software situations. It can efficiently optimise web application speed and cost efficiency, as well as machine learning applications and cloud infrastructure. AIOptimizer uses advanced optimisation and recommendation systems to improve performance and cost. This research study establishes the groundwork for software performance optimisation and emphasises AIOptimizer's potential as a prototype for comparable software solutions.

\section{Future Scope}
\subsection{Test Customization}
    \begin{itemize}
    \setlength{\itemsep}{0pt}
        \item Future objectives for Test Customization include enabling users to define and customise custom test situations and parameters within the software. Users can specify their unique requirements, limitations, and objectives to evaluate the performance and cost optimisation of the software system. This customization may involve defining relevant workloads, performance metrics, and cost factors for their use cases.
        \item By permitting test adaptation, AIOptimizer grows more adaptable to various software systems and allows users to tailor the optimisation process to their particular requirements.
    \end{itemize}
    
    \subsection{Configuration Options}
    \begin{itemize}
    \setlength{\itemsep}{0pt}
        \item AIOptimizer may allow users to customise their algorithms, parameters, and optimisation strategies to their specific preferences and needs.
        \item This may include options to select various reinforcement learning algorithms, exploration-exploitation trade-off options, and hyperparameter fine-tuning. Configuration options enable users to optimise the performance of AIOptimizer based on their domain knowledge and optimisation objectives.
    \end{itemize}
    
    \subsection{Collaboration and Knowledge Sharing}
    \begin{itemize}
    \setlength{\itemsep}{0pt}
        \item Future objectives for Collaboration and Knowledge Sharing include enhancing AIOptimizer with tools that encourage user collaboration and facilitate knowledge sharing.
        \item By exchanging insights, best practices, and optimisation strategies, users can collaborate to optimise software performance and cost. This can be accomplished via chat rooms and group discussions.
    \end{itemize}

\end{document}